          [2001/04/25 1.1 (PWD)]
\documentclass[a4paper,twocolumn]{esapub} 

\usepackage{graphicx,epsf,psfig,epsfig,times,amsmath,amssymb,natbib}
\title{The \textit{INTEGRAL} LMXRB monitoring programme}
\author[1,2]{A. Paizis}
\affil[1]{\textit{INTEGRAL} Science Data Centre, Chemin d'Ecogia 16, 1290 Versoix, Switzerland, Ada.Paizis@obs.unige.ch}
\affil[2]{CNR-IASF, Sezione di Milano, Via Bassini 15, 20133 Milano, Italy}
\author[1]{T.J.-L. Courvoisier}
\author[3]{O. Vilhu}
\affil[3]{Observatory, P.O.Box 14, T\"ahtitorninm\"aki, FI-00014 University of Helsinki, Finland}
\author[1]{M. Chernyakova}
\author[3]{T. Tikkanen}
\author[4]{A. Bazzano}
\affil[4]{CNR-IASF, Sezione di Roma, Via del Fosso del Cavaliere 100, 00133 Roma, Italy}
\author[5]{V. Beckmann}
\affil[5]{NASA Goddard Space Flight Center, Code 661, Greenbelt, MD 20771, USA}
\author[6]{J. Chenevez}
\affil[6]{Danish Space Research Institute Juliane Maries Vej 30,
  Copenhagen, Denmark}
\author[4]{M. Cocchi}
\author[1]{K. Ebisawa}
\author[7]{R. Farinelli}
\affil[7]{Dipartimento di Fisica, Universit\`{a} di Ferrara, Via Paradiso 12, 44100 Ferrara, Italy}
\author[7]{F. Frontera}
\author[8]{A. Gimenez}
\affil[8]{Instituto Nacional de Tecnica Aerospacial, Carretera de Ajalvir 4, Torrejon de Ardoz , Madrid, Spain}
\author[9]{P. Goldoni}
\affil[9]{CEA Saclay, DSM/DAPNIA/SAp, 91191 Gif sur Yvette, France}
\author[3]{D. Hannikainen}
\author[10]{E. Kuulkers}
\affil[10]{Research and Scientific Support Department of ESA, ESTEC,  Postbus 299, 2200 AG Noordwijk, The Netherlands}
\author[6]{N. Lund}
\author[10]{T. Oosterbroek}
\author[11]{S. Piraino}
\affil[11]{CNR-IASF, Sezione di Palermo, Via Ugo La Malfa 153, 90146 Palermo, Italy}
\author[9,1]{J. Rodriguez}
\author[11]{A. Santangelo}
\author[1]{R. Walter}
\author[12]{A. A. Zdziarski}
\affil[12]{N. Copernicus Astronomical Ctr., Bartycka 18, 00716 Warsaw,
  Poland}
\author[1]{J. A. Zurita Heras}

\begin{document}


\maketitle

\begin{abstract}
Our collaboration is responsible for the study of a sample of 72 low
mass X-ray binaries (LMXRB) using the \textit{INTEGRAL} Core Programme data. In this
paper we describe the monitoring programme we have started and the
current variability and spectral results on a
sample of 8 persistently bright LMXRBs hosting a neutron star (Z and
Atoll sources). Current results show that among our sample of sources there seems  to be no important difference in the variability of Z
sources with respect to Atolls and the
first colour-colour and hardness intensity diagrams built in the "traditional" energy bands display
the expected patterns. \\
Z sources seem to be harder than the bright Atolls of our sample (above 20 keV) and present no evident cut-off until about 50
keV. A hint of a non-thermal hard tail is seen in
Sco X-1 with ISGRI and SPI, similarly to what was previously detected by D'Amico et al. (2001) with \textit{RXTE}.\\
These results, even if preliminary, show the importance of such a programme
and the potential underlying it to understand these
sources as a class.
\end{abstract}

\section{Introduction}

The International Gamma-Ray Astrophysics Laboratory,
\textit{INTEGRAL} (Winkler et al. 2003), has been launched in October 2002. Since then, it has been providing a large
amount of interesting data. The combination of  two wide
field of view (FOV) instruments, the imager IBIS (15\,keV -- 10\,MeV,
$29^{\circ}\times 29^{\circ}$ partially coded FOV, Ubertini et al. 2003) and the spectrometer SPI (20\,keV -- 8\,MeV, $35^{\circ}\times
35^{\circ}$ partially coded hexagonal FOV, Vedrenne et al. 2003) 
coaligned with the JEM--X
(Lund et al. 2003) and OMC (Mas-Hesse et al. 2003) monitors, allows large
areas of the sky to be observed and monitored simultaneously in a
wide energy range from the optical to the $\gamma$-ray domain.
Such a capability is fully exploited during the \textit{INTEGRAL} Core
Programme (Winkler et al. 2003b), a series of successive scans of the Galactic Plane (GPS)
and Galactic Centre (GCDE), which is regularly producing large amounts
of data, in particular on persistently bright sources.\\
Our collaboration is responsible for the monitoring of a sample of low mass
X-ray binaries (LMXRBs). In this paper we describe the current results of this
programme, showing the importance of this study and the potential
underlying such a long term monitoring. \\
Section 2 of this paper describes the aim of the monitoring programme
with a basic overview of the source characteristics. Section 3
contains the data reduction and analysis description while in Section
4 the current results are given. In the last Section we present our conclusions
and future plans.

\section{The monitoring programme}
\subsection{The sources of our sample}
The nature of the sources in our list is very rich, containing black hole
(BH) as well as weakly magnetised neutron star (NS) binaries with very
different variability.\\
Monitoring these sources through the years will give an overview of
the hard energy ($>$ 10 keV) behaviour of the Galactic Plane and
Centre LMXRBs as a class: outburst frequency, variability level, type
I X-ray burst frequency, persitent emission etc, all in the poorly studied hard
energy domain.\\
Among all the sources of our sample, in this paper we focus on the 8
persistently bright LMXRBs listed in Table 1. They all host a neutron star.
\begin{table}
  \begin{center}
    \caption{Bright persistent NS LMXRBs regularly monitored by \textit{INTEGRAL}.
  \emph{Type:} A=Atoll, B=bursting, Z=Z source, ADC=Accretion Disc Corona.}\vspace{1em}
    \renewcommand{\arraystretch}{1.2}
    \begin{tabular}[h]{lrcc}
      \hline
      Source Name & l      &  b      &  Type \\
      \hline
Sco X$-$1&359.09 & 23.78& Z  \\
Cyg X$-$2& 87.33 & -11.32 & ZB \\
GX 5$-$1&  5.08 &  -1.02 &    Z  \\
GX 17$+$2& 16.43 &   1.28 &   ZB \\
GX 3$+$1&  2.29 &   0.79 &   AB\\
GX 9$+$9&  8.51 &   9.04 &  A   \\
GX 9$+$1&  9.08 &   1.15 &    A \\
4U1822$-$371& 356.85 & -11.29 & ADC \\ 
\hline \\
      \end{tabular}
    \label{tab:table}
  \end{center}
\end{table}
\subsection{Weakly magnetised neutron star binaries}
The current classification of NS LMXRBs is based on the pattern displayed
by individual sources in the X-ray colour-colour (CC) and
hardness-intensity (HI) diagrams (Hasinger \& van der Klis, 1989 and
van der Klis, 1995). It comprises the so-called Z sources (that
display a "Z" pattern in the diagrams) and the Atoll sources (that
display an upwardly curved branch in the diagrams). \\
More recent studies (Muno et al. 2002, Gierli\'nski \& Done 2002
and Done \& Gierli\'nski 2003)
suggest that the clear Z/Atoll distinction in the CC 
diagram is an artifact due to incomplete sampling: Atoll sources, if
observed long enough, \emph{do} exhibit a Z shape in the CC as
well. Many differences, however, remain: Atoll sources have weaker
magnetic fields ($<$$10^9$--$10^{10}$\,G versus $\sim$10$^9$--10$^{10}$\,G of
Z sources),  are generally fainter ($0.01$--$0.3 L_{Edd}$
versus $\sim L_{Edd}$), can exhibit harder spectra, trace out the Z shape on longer time
scales than typical Z sources and have a different correlated
timing behaviour along with the position on the Z.
Thus the distinction, at least from a practical point of view, still
makes sense.\\
Our sample of sources includes both Z and Atoll sources. Thanks to
the \textit{INTEGRAL} monitoring programme we (will) have a long term
coverage of all these sources. For the first time they will be studied
in a regular and unbiased way in the energy band in which they are
poorly known, hard X/$\gamma$ rays, where they display an interesting
behaviour.\\
Thermal comptonisation is dominant both in soft and hard spectral states of
NS LMXRBs and in most cases it is a good representation of
the spectra below 20 keV.
In this range, LMXRB with a weakly magnetised NS (i.e. non pulsating) can be well
described by two competing models. On one side, there is the so-called
"western" model in which the spectrum is composed of the sum of unsaturated Comptonised spectrum (produced by an inner disc corona) plus
 a blackbody originating  from a region close to the neutron star surface or from the boundary layer
between the disc and the neutron star  (White et al. 1986, 1988). On the other hand, in the "eastern" model  the
spectrum consists of the sum of an optically thick multi-temperature
disc-model (locally emitting like a pure blackbody) plus a comptonised
blackbody again originating from the neutron star or boundary layers
(Mitsuda et al. 1984; Mitsuda et al. 1989).\\
Hard X-ray components extending up to several hundred keV have been revealed in
about 20 NS LMXRBs of the Atoll class (Di Salvo \& Stella 2002
and references therein). 
In these systems a power-law like component (with photon index $\Gamma
\sim 1.5-2.5$) is followed by an exponential cutoff between $\sim$20 and many tens of  keV. This is explained in terms of unsaturated
thermal Comptonisation. But there are cases in which no evidence for a
cutoff is found up to 100-200 keV. This is the so-called "hard
state" of Atoll sources and occurs especially in the lower luminosity
systems (note that the Atolls of our sample are among the brightest). On the other hand, broad band
studies have shown that also many Z sources display a variable
hard power-law ($\Gamma \sim 1.9-3.3$) component dominating the spectra
above $\sim 30 \rm \, keV$. \\
The origin of these hard tails is still debated. Radio observations of
some Z sources (Fender \& Hendry 2000) seem to show a general trend
in which the highest radio fluxes (thought to be originating in jets)
are associated to the hardest state of the sources. This could mean
that the non thermal, high-energy electrons responsible for the hard
tails in Z sources could be accelerated in jets (Di Salvo et
al. 2000).
\subsection{Aim of the programme}
The long term X-ray variability of LMXRBs has been extensively studied in the
2--12\,keV band with the \textit{RXTE} All Sky Monitor. At higher
energies, the information gathered from the sources has been obtained
mainly via dedicated pointings (\textit{RXTE}, \textit{BeppoSAX} etc). \\
The combination of regular monitoring in the hard X-rays
and $\gamma$-rays has not been done before and this is where
\textit{INTEGRAL} will give a major contribution to understand the
behaviour of bright LMXRBs from 5 keV up to $\sim
200 \, \rm keV$ (see also Paizis et al. 2003).\\
By focusing on the sources presented in this paper, we intend to try to understand the differences
between Z and Atoll sources via their (less explored) high-energy
emission (hard tails, CC diagrams, different variability, etc).\\
In collaboration with the accreting pulsars collaboration, we intend to put the extracted light-curves and hardness intensity diagrams on
the web\footnote{The results will thus be publically available,
similarly to what the \textit{RXTE}/All Sky Monitor has been doing in the softer X-ray band.}.
In this way, the high-energy history of these sources will be easily
accessible, enabling also multiwavelength comparisons. In
this respect, our collaboration has coordinated \textit{RXTE} observations to the
\textit{INTEGRAL} ones in order to have a better coverage of the soft
X-ray domain \footnote{In fact, due to the GPS and GCDE dithering patterns, the sources are very
often only in the partially coded FOV of JEM-X and sometimes
even not covered at all.}. Similarily,
we have access to Radio (RATAN/MOST/VLA) and Optical (La Palma, La
Silla) telescopes.
\section{Data reduction and analysis}
\textit{INTEGRAL} performs a GPS about every 12 days and the GCDEs are
performed according to the Galactic Centre visibility. At the time of
writing the first year of core programme has been completed and we have
analysed all the data from revolution 26 (January 2003) until
revolution 142 (December 2003) for a total of 3078 science windows
corresponding to about 5.7 Msec. \\
Version 3.0 of ISDC's (Courvoisier et al. 2003) Off-line Science Analysis (OSA) software 
 has been used for analysing the data.\\
 Given the type of sources involved (rather steep spectra with about hundred of mCrabs in the 2--10\,keV band, Liu
et al. 2001) and the pointing exposures of about 2000 s, for the
analysis we have chosen JEM--X for soft photons (5--20\,keV) and the low energy IBIS detector, ISGRI
(Lebrun et al. 2003) for harder photons (20--200\,keV). We have not
used PICsIT, the hard photon IBIS detector (Di Cocco et al. 2003), as its 
peak sensitivity is above 200 keV, where LMXRBs have fluxes below PICsIT detectability.
The spectrometer SPI and the imager ISGRI  have been used to extract the hard energy spectra
(20-200\,keV) averaged on longer time scales.  To increase the signal
 to noise ratio in the extracted ISGRI spectra, we used an alternative
 method i.e. we first combined  data of different pointings in one
 single mosaic (weighted combination of the images) and then extracted fluxes in several 
 energy bands.
\begin{figure*}
\centering
\includegraphics[width=1.0\linewidth]{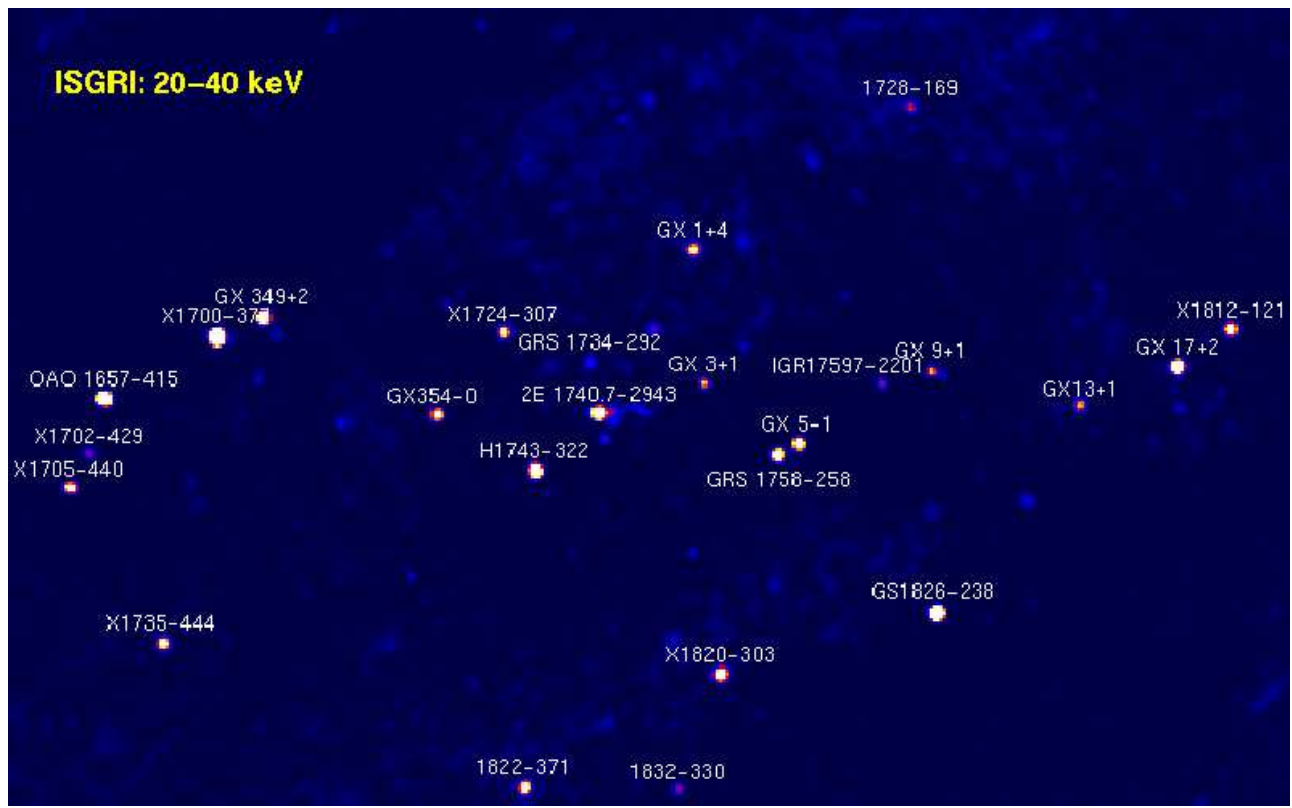}
\caption{IBIS/ISGRI (20--40 keV) mosaic of the Galactic Centre (5.7
  Msec, about $20^{\circ}$x$40^{\circ}$  centred on the Galactic Centre). Only a few sources
  are labelled for clarity. Most of the sources of this paper are
  visible in the image (GX9+9 is labelled here as 1728-169). \label{fig:ima}}
\end{figure*}
\section{Results}
\begin{figure*}
\centerline{
\begin{tabular}{cc}
\psfig{file=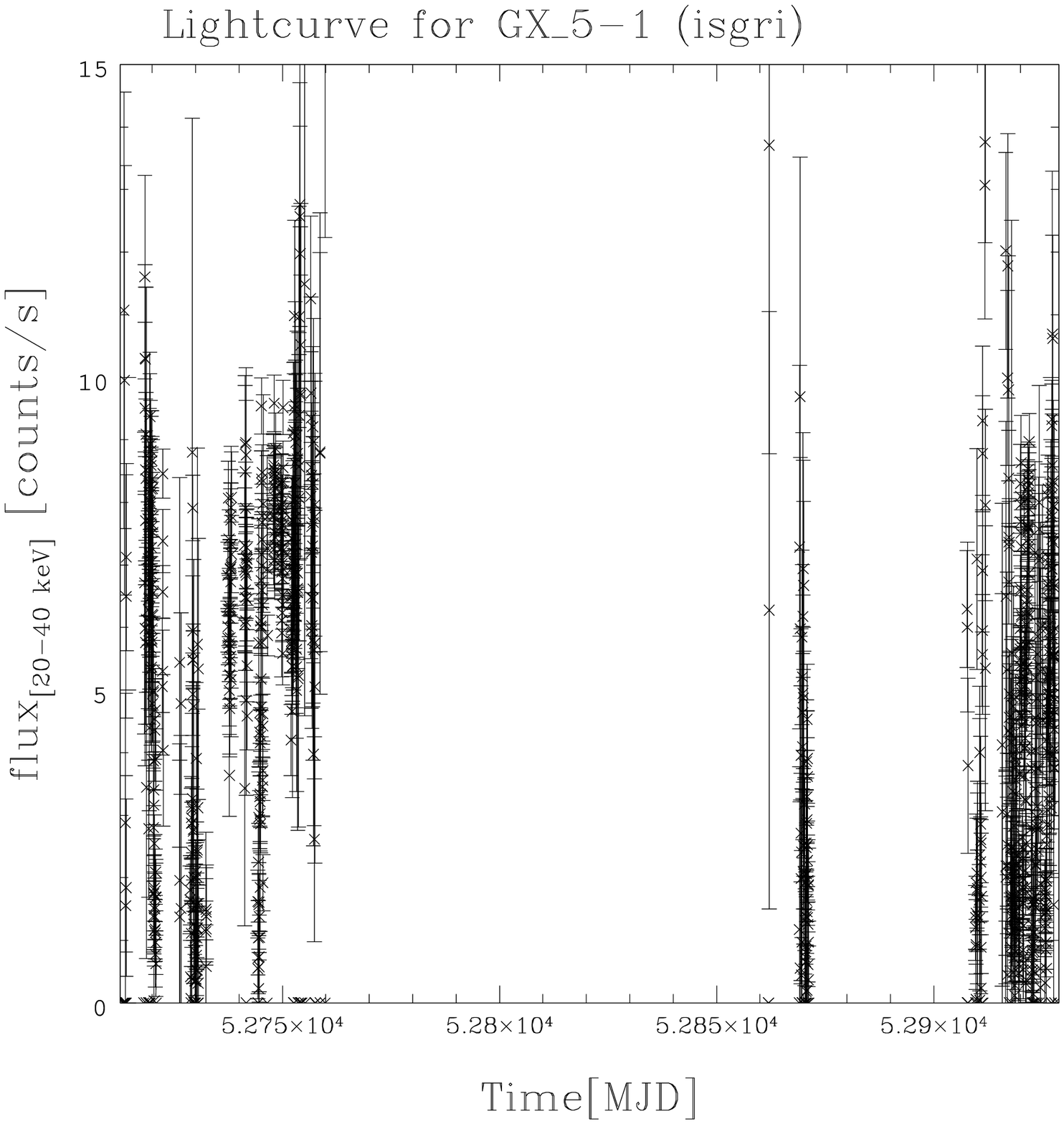, width=0.5\linewidth} &
\psfig{file=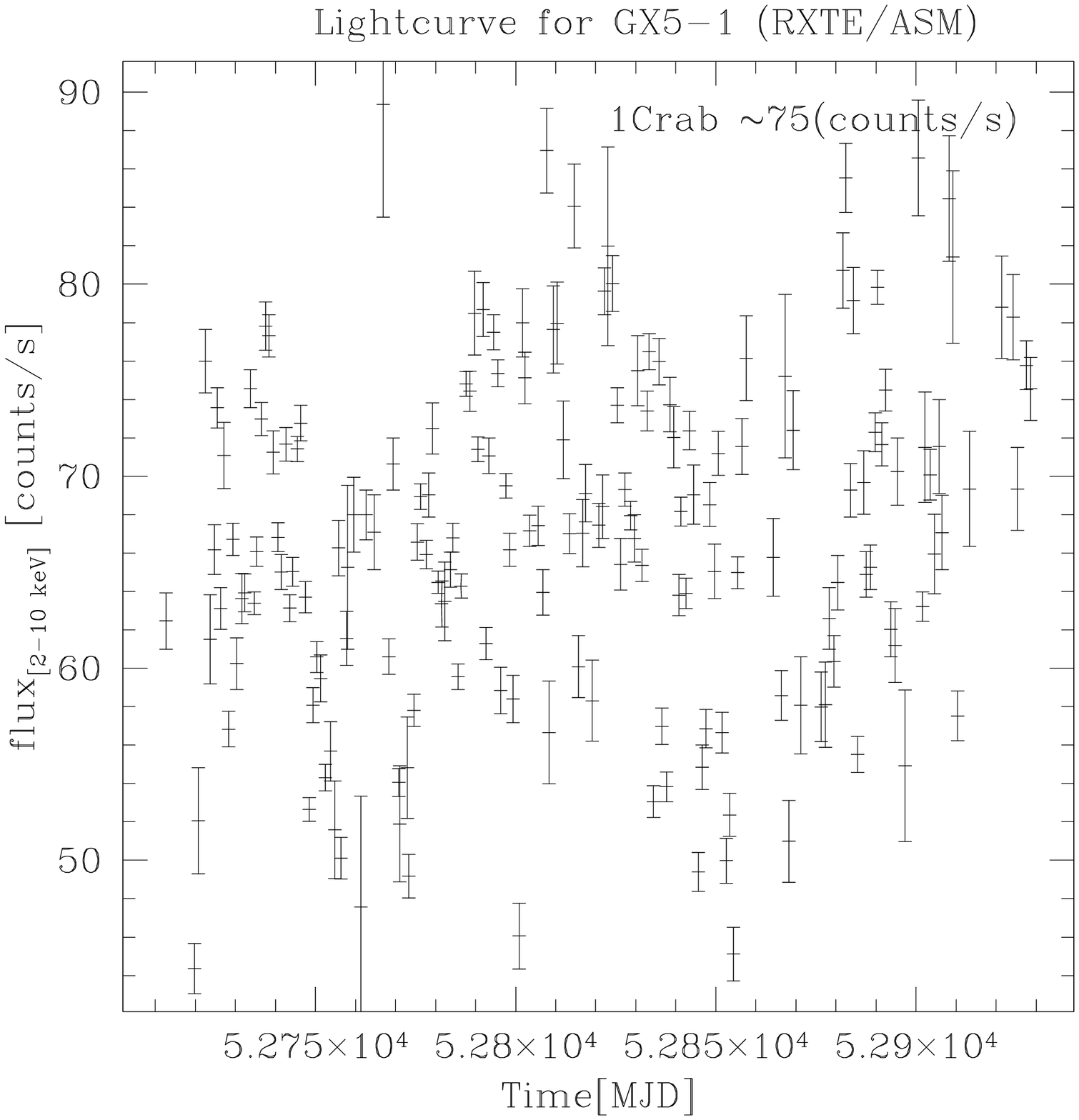, width=0.5\linewidth} 
\end{tabular}}
\caption{GX5-1 light-curves from March 2003 until October
  2003. \emph{Left panel}: ISGRI
  results in the 20-40 keV band (about 2000 sec time-bin). In this
  energy band 1Crab corresponds to about 100 counts/sec.
  \emph{Right panel}:  quick-look results provided by the \textit{RXTE}/ASM
  team in the 2-10 keV band (1 day time-bin).  \label{fig:months}}
\end{figure*}

\begin{figure*}
\centerline{
\begin{tabular}{cc}
\psfig{file=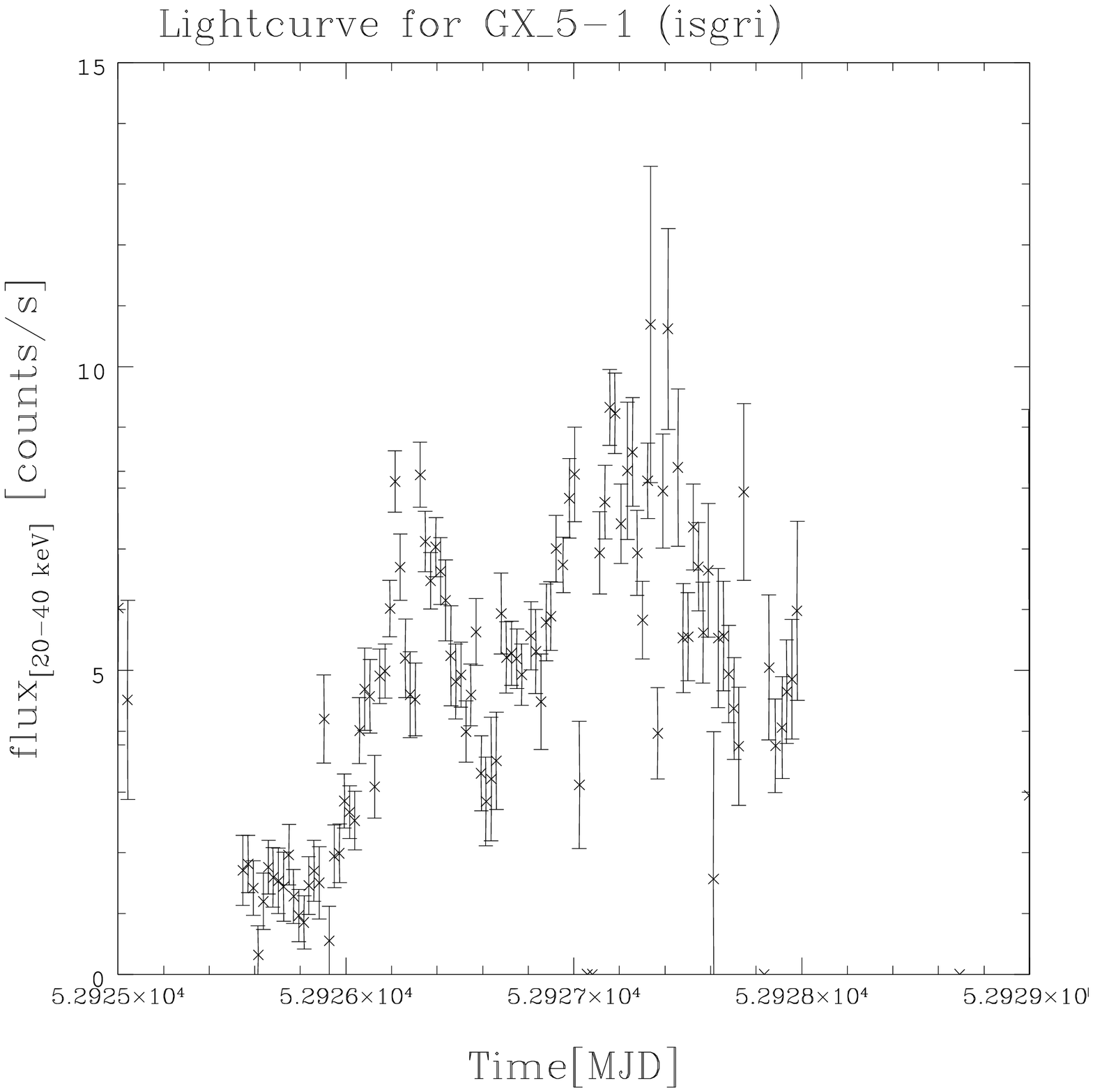, width=0.5\linewidth} &
\psfig{file=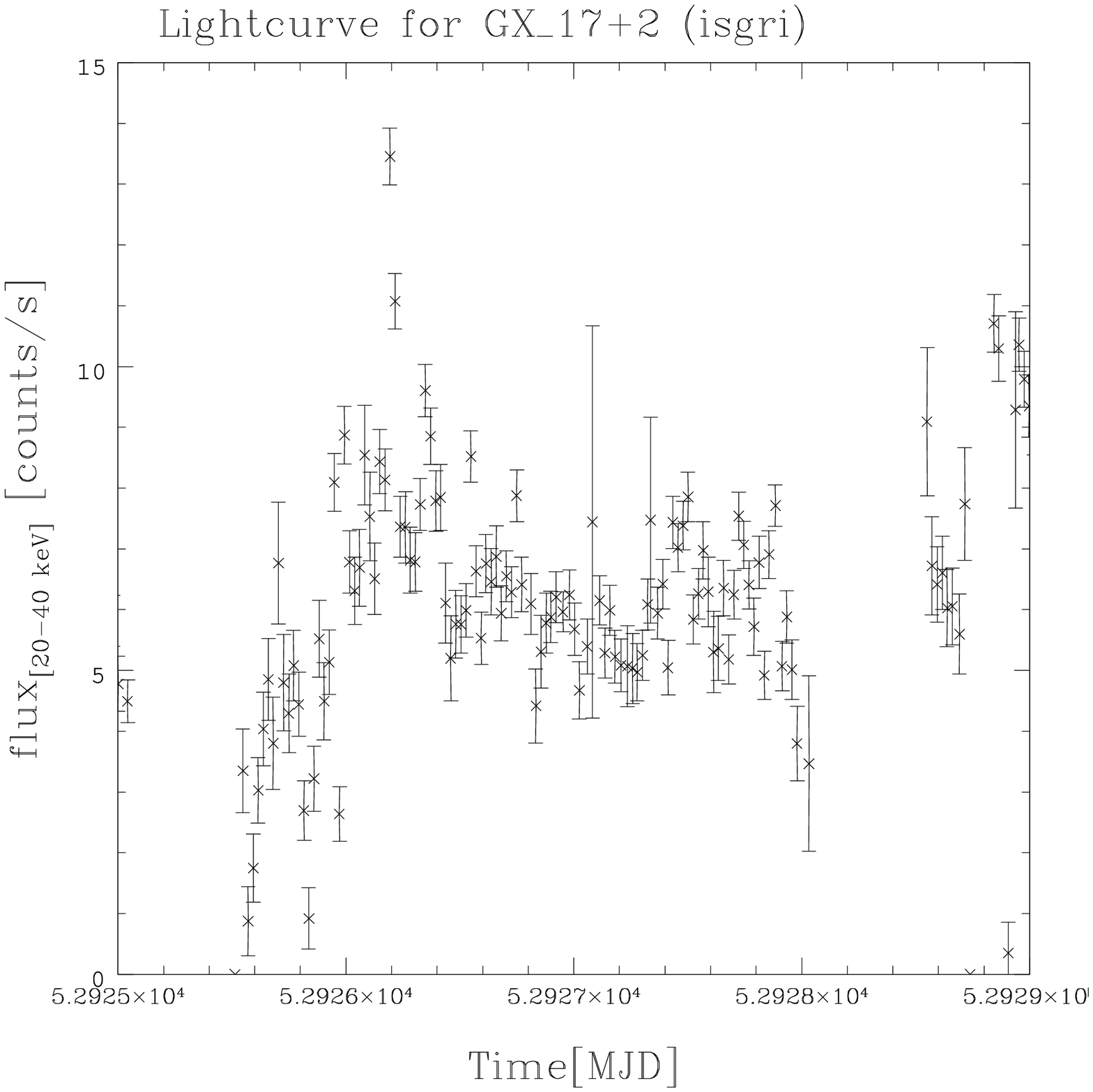, width=0.5\linewidth}
\end{tabular}}
\caption{5-day ISGRI light-curve for GX5-1 (\emph{left panel}) and
  GX17+2 (\emph{right panel})  \label{fig:days}}
\end{figure*}
In this section we go through the main results that we have obtained
during our monitoring programme up to now. They can be mainly split into two
parts: the variability study of our sources and the (average) spectral 
study. \\
Fig.~\ref{fig:ima} shows the distribution in the Galaxy of the sample of sources
studied in this paper (with the exception of Sco X-1 and Cyg
X-2\footnote{Sco X-1 and Cyg X-2 are far away from the galactic
  plane and actually are never covered by JEM-X during the scans. This
  makes our simultaneous \textit{RXTE}  coverage even more important.}). The
total exposure time is about 5.7 Msec while the final exposure time
per source depends on its position with respect to the Galactic
Centre: the closer to the Centre the higher the exposure. 
\subsection{Variability study}
\subsubsection{Light-curves}
\begin{figure}
\centering
\includegraphics[width=1.0\linewidth]{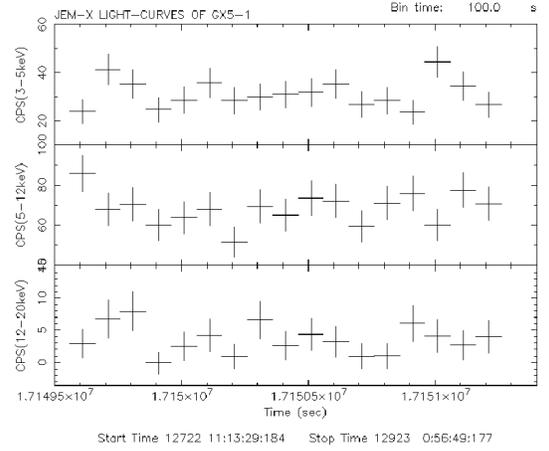}
\caption{Single pointing JEM-X light-curve of GX5-1 (100 sec bins) in
  three different energy bands. \label{fig:jemx}}
\end{figure}

\begin{figure}
\centering
\includegraphics[width=1.0\linewidth]{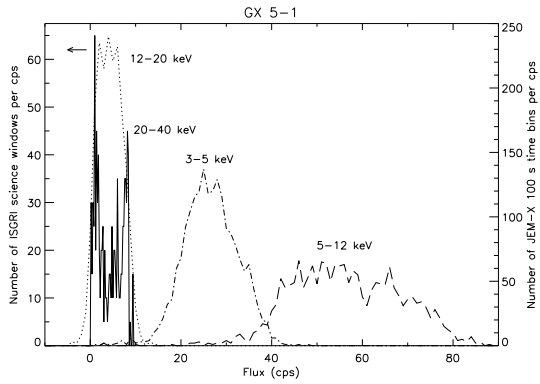}
\caption{Count rate distribution for GX5-1. The solid line is the 20-40 keV ISGRI
  distribution while the remaining three curves are the distributions
  in the three different JEM-X
  bands used in our analysis. \label{fig:CRDZ}}
\end{figure}

\begin{figure}
\centering
\includegraphics[width=1.0\linewidth]{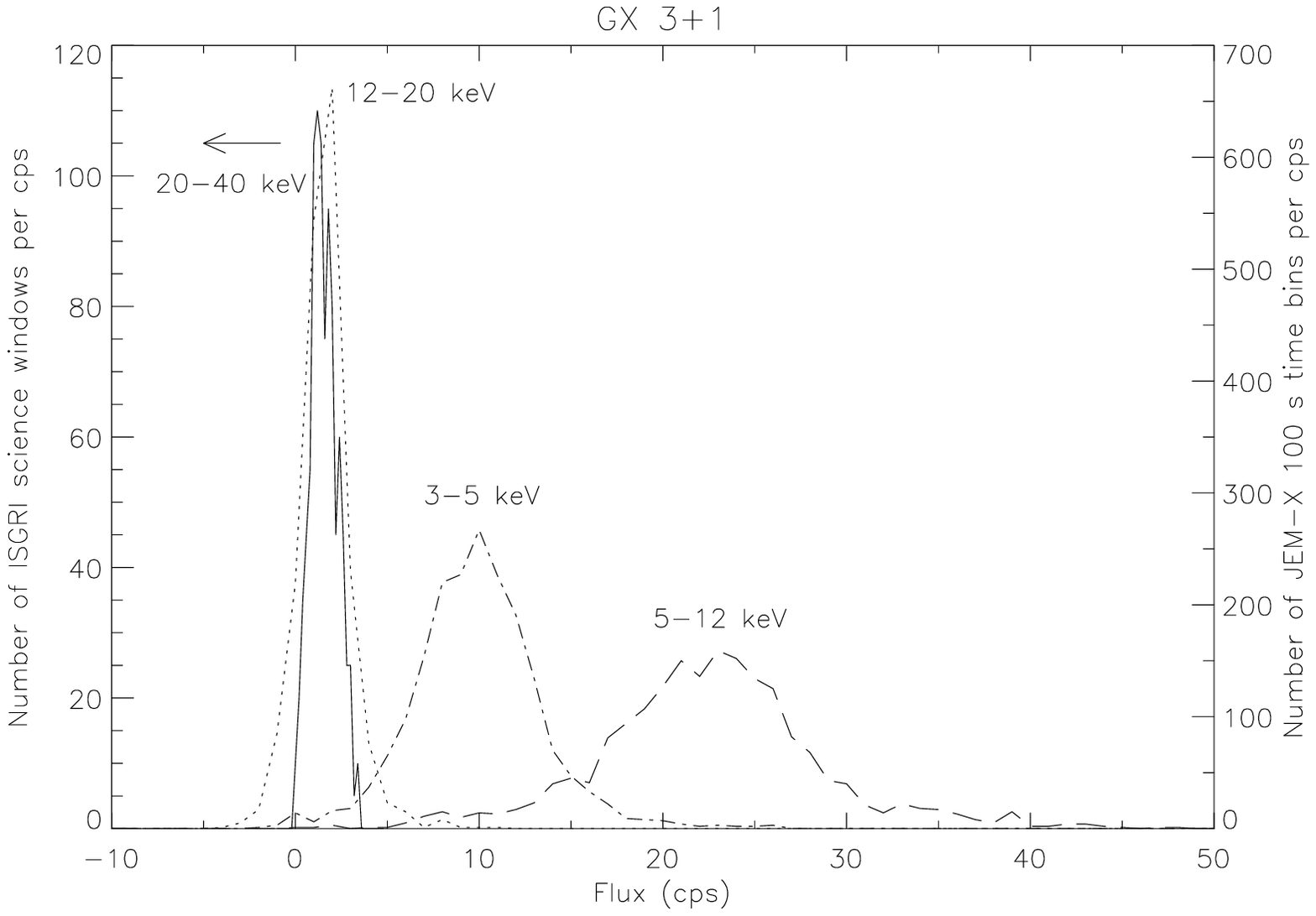}
\caption{Count rate distribution for GX3+1 in the same energy bands of Fig.~\ref{fig:CRDZ}\label{fig:CRDA}}
\end{figure}

As already stated, one of the main aims of the monitoring programme is
to extract light-curves for the sources in different energy
bands. The richness of the light-curve depends on the position of
the source in the sky. Fig.~\ref{fig:months} (left panel) is an example of an ISGRI
light-curve for GX5-1 (quite close to the Centre of the Galaxy). The
points with larger error bars correspond to pointings where the source
was more off-axis. \\
For comparison the right panel shows the \textit{RXTE}/ASM light-curve for the
same period (7 months). The ASM coverage is of course more extensive (\textit{INTEGRAL}
has no  all sky monitor programme) but with \textit{INTEGRAL}/ISGRI we will have a
coverage of the sources in the harder energy
bands, completing the ASM. A hard-energy (20 keV -- 1 MeV) sky survey has been performed
by the BATSE mission aboard \textit{CGRO} covering the period from
April 1991 until June 2000 (Shaw et al. 2004). The \textit{INTEGRAL}
high-energy survey will achieve a much better angular resolution and
sensitivity\footnote{Integrating over the full nine year database of
  BATSE observations and over 7 energy channels (25 -- 160 keV), the
  5$\sigma$ sensitivity to a persisitent source is $\sim$2mCrab while the
  angular resolution achieved with BATSE is about half a degree.}.\\
Fig.~\ref{fig:days} shows a zoom in a 5-day ISGRI
light-curve of GX5-1 and GX17+2. \\
 Light-curves similar to Fig.~\ref{fig:months} and Fig.~\ref{fig:days}
 are being extracted also with JEM-X. In this case, due
to the smaller size of the FOV, the resulting data set is smaller than
for ISGRI. On the other hand, JEM-X covers a softer part of the X-ray
spectrum and thus has more statistics. This allows the extraction of smaller
time-bin light-curves like the one showed in Fig.~\ref{fig:jemx} where 1
pointing (science window) is further sampled into  100 second bins.
Starting from this data base of light-curves (1 pointing bin for ISGRI
and 100 seconds bin for JEM-X) we have produced, per source, the count
rate distributions in different energy bands. The distributions for GX5-1 (Z
source) and GX3+1 (Atoll) are shown in Fig.~\ref{fig:CRDZ} and
Fig.~\ref{fig:CRDA} respectively, while Table 2
summarises the results we obtain for all the Z and Atoll sources of our
sample. 
\begin{table}
  \begin{center}
    \caption{Variability properties of the sources. \emph{J\_Mean}: mean counts/sec in the 5--12 keV JEM-X
      band. \emph{J\_Var}: standard deviation of the distribution
    normalised to the mean in  the 5--12 keV JEM-X
      band in \%. \emph{I\_Mean}: mean counts/sec in
    the 20--40 keV ISGRI band. \emph{I\_Var}: standard deviation of the distribution
    normalised to the mean in  the 20--40 keV ISGRI
      band in \%.   }\vspace{1em}
    \renewcommand{\arraystretch}{1.2}
    \begin{tabular}[h]{ccccc}
      \hline
      Source & J\_Mean & J\_Var.& I\_Mean & I\_Var.  \\
      \hline
      \hline
      Z  & & & & \\
      \hline
GX 5$-$1& 56.96 & 39\% & 4.46 & 64\% \\
GX 17$+$2& 40.30 & 35\% &5.82 & 42\% \\
Sco X$-$1& - & - & 78.20 & 32\% \\
\hline 
\hline
      ATOLL & & & & \\
      \hline
GX 3$+$1& 23.04 & 42\% & 1.65 & 44\%\\
GX 9$+$9& 14.77 & 36\% & 1.35 & 42\%  \\
GX 9$+$1& 32.41 & 41\% & 1.72 & 41\%\\
\hline 
\hline
      ADC & & & & \\
      \hline
1822$-$371&1.34 & 87\% & 3.35 & 26\% \\ 
\hline
      \end{tabular}
  \end{center}
\end{table}
Only the pointings where both ISGRI and JEM-X data were
available have been considered. The spread of the distributions is
most likely mainly due to the source variability: the poissonian spread
accounts for a few \% and the vignetting factor (more difficult to
quantify) seems to play a minor role since the dependency of the count
rate on the off-axis angle has been studied and shows no evident trend
for the different sources.
What can be seen from the current data set is that Z sources are
brighter than Atoll sources (as expected) and there seems to be no
important difference in the long-term ($>$ 100 sec) variability of these sources as a class. Apart
from GX5-1, that shows an important variability increase when moving
from soft to hard range, the remaining Z and Atoll sources do not seem to have
evident differences\footnote{The soft (JEM--X) Sco X-1 varibility is missing because JEM-X FOV is too small
to cover it during the scans. Our \textit{RXTE} observations of this source
will provide a simultaneous soft (\textit{RXTE}) hard (\textit{INTEGRAL})
variability. The same holds for Cyg X-2 (not covered by JEM-X) for
which we currently have too few ISGRI points for the count rate
distribution (the source is covered only in the GPS and not on the
GCDE).}.\\
On the contrary, the ADC source 4U1822-371  displays a very high flux
change in the softer energy range. This result is most
likely due to the nature of this source that is known (e.g. Parmar et
al. 2000) to display deep variations in the form of regular dips
and coronal partial eclipses (hence the name of accretion disc
corona source).


\subsubsection{Colour-colour and hardness-intensity diagrams}
\begin{figure}
\centering
\includegraphics[width=1.0\linewidth]{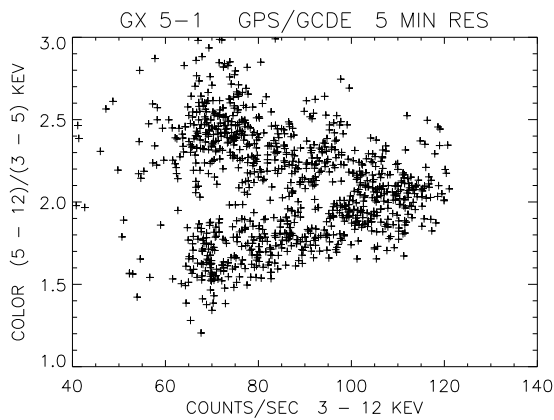}
\caption{Hardness intensity diagram for GX5-1 obtained with JEM-X (one
  year data, 5 minute bins). \label{fig:CC}}
\end{figure}
\begin{figure}
\centering
\includegraphics[width=1.0\linewidth]{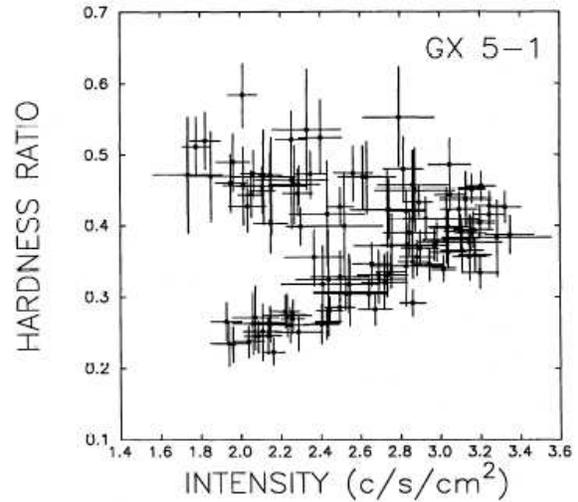}
\caption{Hardness intensity diagram for GX5-1 obtained with \textit{Ginga}-All Sky Monitor (three years
  data). The \textit{Ginga}-ASM hardness ratio is built with counts
  in the 6--20\,keV/1--6\,keV bands while the intensity is the overall
  count rate in the 1--20\,keV band (van der Klis et al. 1991). \label{fig:CCGINGA}}
\end{figure}

The changes in X-ray spectra of Z and Atoll sources are very subtle
and not easy to spot and describe with proper model fitting. 
Alternative tools are often used to study the spectral variability of
these sources and are the already mentioned colour-colour and hardness-intensity diagrams.\\
An attempt is made to build these CC and HI diagrams in the same
energy bands (i.e. defining the same colours) so that time-distant
observations, often performed with different instruments, can still be
compared to have a long term view of the source variability. But this
is not always possible and normally it is difficult to derive a direct
quantitative comparison of CC and HIDs produced with different X-ray
detectors.  In this respect, the \textit{INTEGRAL} monitoring has the main
advantage that the sources (Z and Atolls) will be observed over the years with the
same instrumentation and long-term comparisons will be possible.\\
 Besides, based on the huge data base that will be populated with time, it will
be possible to search for differences and/or similarities of these
sources in new, \textit{INTEGRAL} defined, hard X-ray colour colour diagrams. \\
The systematic study of CC diagrams in different  colours
is still on-going and is very closely related to the status of the
calibration of \textit{INTEGRAL} instruments. Nevertheless, the first results
based on the traditional colour definition seem to show that we indeed
obtain the expected pattern for the sources. Fig.~\ref{fig:CC} shows the JEM-X
HID for GX5-1 to be compared to  Fig.~\ref{fig:CCGINGA}, the
\textit{Ginga}-All Sky Monitor HID (van der Klis et
al. 1991). The
horizontal and normal branches of the Z pattern are clearly visible.
\subsection{Spectral study}
\begin{figure}
\centering
\includegraphics[width=1.0\linewidth]{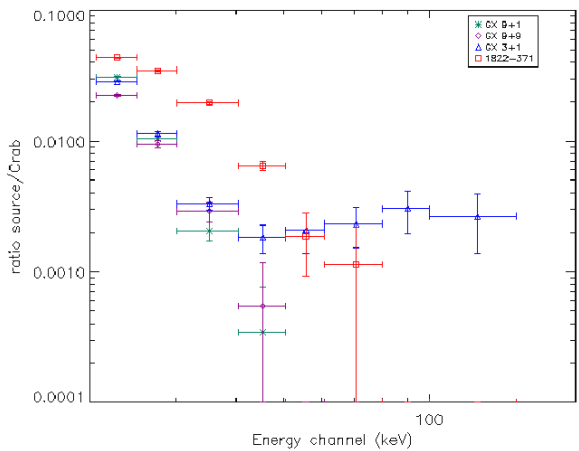}
\caption{ISGRI spectra extracted from a mosaic of 1 year of core
  programme data. Atoll sources plus the ADC source. \label{fig:SPEA}}
\end{figure}

\begin{figure}
\centering
\includegraphics[width=1.0\linewidth]{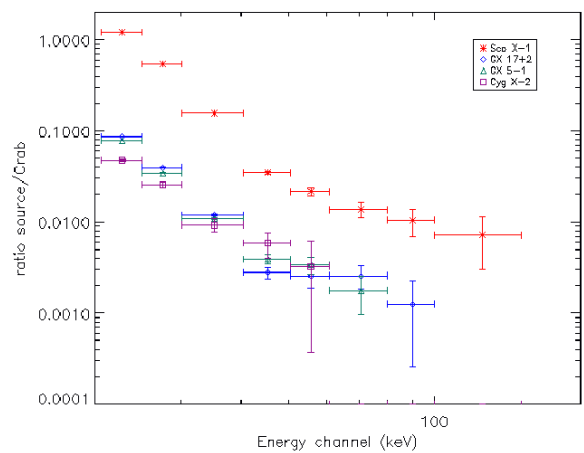}
\caption{ISGRI spectra extracted from a mosaic of 1 year of core
  programme data. Z sources. The effective exposure of each
  source depends on its position (see text). The GX5-1 
  (close to the Centre of the Galaxy) spectrum exposure  is about
  1.6 Msec while for Sco X-1 (higher latitude) it is about 260 ksec. \label{fig:SPEZ}}
\end{figure}

Starting from the ISGRI imaging results, we have built several mosaics,
one per energy band, and then extracted the spectra for each source.
Fig.~\ref{fig:SPEA} and~\ref{fig:SPEZ} show the resulting ISGRI (average) spectra. The
spectra have been normalised to the Crab spectrum (extracted in the
same way): a zero slope in the graphs means a source as hard as Crab
(i.e effective photon index of 2) while a positive slope means a
source softer than Crab. \\
In Fig.~\ref{fig:SPEA} the 3 Atoll sources (GX9+1, GX9+9 and GX3+1) have a similar soft spectrum until about 50 keV. Above
50 keV, GX3+1 shows a hardening comparable to the hardness of the Crab
with a $4.5 \sigma$ significance in the last 3 bins. Such hardening can be described by a
comptonised black-body component and (with the current systematics) no additional power-law
component (hard tail) is needed. It is also important to note that the
source brightness above 60 keV (where the hardening is more evident)
is around a few mCrab i.e. comparable to the background fluctuations
at these energies (Bodaghee et al. 2004). It is currently difficult
to disentangle among source and background contribution and a complete
calibration of the instrument is needed to derive more firm conclusions.\\
\begin{figure}
\centering
\includegraphics[width=1.0\linewidth]{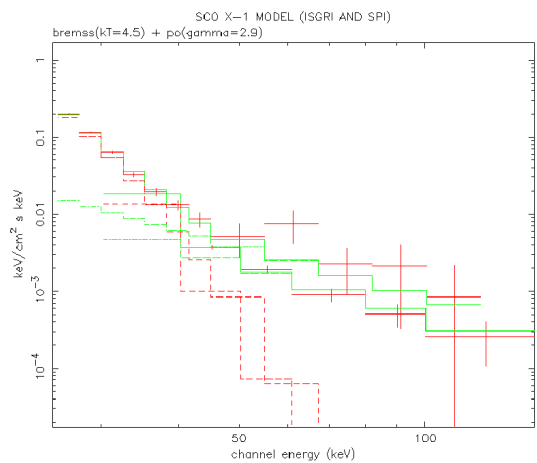}
\caption{Energy ISGRI and SPI spectra of Sco X-1. The SPI data
  are the ones with larger error bars. The dotted lines are the
  bremsstrahlung and power-law models (for the 2 spectra). The solid line is the total model.\label{fig:SCO}}
\end{figure}
Fig.~\ref{fig:SPEZ} shows the spectra of the Z sources of
our sample. Z sources are brighter than Atolls (as expected) and seem
also to be harder with no evident cut-off until about 50 keV. Sco X-1
 shows  a hardening above 50 keV. In this case the
hardening is more significant than for GX3+1 (10 $\sigma$ detection in the 55 keV centred
bin and 5.3 $\sigma$ detection in the 70 keV centred bin) and starts well
above the aforementioned background limit. Triggered by this, we have
performed a deeper spectral study. Fig.~\ref{fig:SCO} shows
 ISGRI and SPI spectra of Sco X-1. We fitted the data with the
 best fit model  that D'Amico et al. (2001) used  to describe
 the non-thermal hard tail detected in  Sco X-1 with \textit{RXTE}/HEXTE instrument. We get
 comparable results for the bremsstrahlung component (temperature of
 about 4.5 keV) and a slightly steeper power-law slope (2.9 instead of their
 maximum 2.37). In our case the slope of the power-law component is
 difficult to determine accurately given that it strongly depends on
 the softer bremsstrahlung component which, in turn, depends on the
 softer ($<$ 20 keV) part of the spectrum  currently missing (the
 source is not covered by JEM-X and the simultaneous \textit{RXTE} data have
 not been triggered yet). Our detection confirms the non-thermal hard
 tail detection of \textit{RXTE}/HEXTE by  D'Amico et al. (2001) . Nevertheless, such a result
 should be taken with caution: ISGRI calibration is not optimised
 yet and a hard tail as the one we detect on a $\sim$260 ksec averaged spectrum would
 mean that the tail is either steadily there (which does not seem the
 case from previous observations on Sco X-1) or indeed is variable but
 very strong when present. The latter could be the case and we will
 extract spectra at a (few) science window(s) level to have the
 answer.
\section{Conclusions}
We have analysed about 1 year of \textit{INTEGRAL} Core Programme data and built
a LMXRB data base that will be made publicly available via the web.\\
In our monitoring progam we plan to study in a systematic way the high-energy emission of a sample of 72 LMXRBs. Among these are 8 persistently bright neutron
star LMXRBs (4 Z sources, 3 Atolls and 1 ADC source).  In this paper
we have shown the current results from this sample of 8 LMXRBs (all
hosting a weakly magnetised neutron star). \\
The \emph{variability study} (light-curves and CC-HI diagrams) has shown that
the \textit{INTEGRAL} core programme coverage is enough to study the high-energy history and evolution of
the sources. 
Z sources are brighter than Atolls (as expected) and, with the current data set, there seems not to be an
important difference in the variability of the sources as a class. The
CC-HI diagrams built in the "traditional" energy bands already display
the expected patterns which is an encouraging result for exploring new,
\textit{INTEGRAL} defined, diagram energy bands.\\
The \emph{spectral study} of the sources of our sample has shown that Z sources seem to be harder than
Atolls ($>$ 20 keV) and present no evident cut-off until about 50
keV. Atoll sources in general, as previously stated, can be much harder than Z
sources but this is mainly true for the low luminosity ones whereas the Atolls of
our sample are soft high state bright systems.\\
In our averaged ISGRI spectra, a hardening in GX3+1 data is visible (well described by
the traditional comptonised black-body model, i.e. no additional
power-law component needed) and a hint of a non-thermal hard tail in
Sco X-1 ISGRI and SPI data is seen, similarly to what was previously detected by D'Amico et
al. (2001) with \textit{RXTE}.\\
 The hunt for such hard tails in NS LMXRBs is a key goal of our monitoring with \textit{INTEGRAL}.
They add one more piece to a mosaic
 that places neutron star binaries next to black holes, for which non-thermal emission was thought to be a prerogative.\\
In the results presented here, variability and spectral studies have been carried out
separately and the next step will be to merge these two aspects,
extracting only spectra for a given branch in the CC-HI
diagrams, i.e. for well defined spectral states. The coordinated observations will show the presence (or
absence) of multiwavelength emission in the different states.\\
Using all this we are able to build a huge data base that will offer a unique long-term,
regular and energy-wide study of a sample of (intrinsically different)
LMXRBs. We expect this to be a step forward in the understanding of the physics and
geometry of X-ray binaries.



\begin{thebibliography}{}
\bibitem[2004]{arash}
Bodaghee, A., Walter, R., Lund, N., \& Rohlfs, R. 2004, these proceedings
\bibitem[2003]{thierry}
Courvoisier, T.J.-L., Walter, R., Beckmann, V., et al. 2003, \textit{A\&A} 411, L53
\bibitem[2001]{damico}
D'Amico, F., Heindl, W. A., Rothschild, R. E., et al. 2001, \textit{ApJ} 547, 147
\bibitem[2003]{dicocco}
Di Cocco, G., Caroli, E., Celesti, E., et al. 2003, \textit{A\&A} 411, L189
\bibitem[2000]{disalvo2}
Di Salvo, T., Stella, L., Robba, N. R., et al. 2000,  \textit{ApJ} 544, L119
\bibitem[2002]{disalvo}
Di Salvo, T., \& Stella, L. 2002, Proc. of the XXXVIIth Rencontres de
Moriond, \textit{The Gamma-Ray Universe}, Ed. A. Goldwurm, Doris N. Neumann and Jean Tran Thanh Van, 67, astro-ph/0207219
\bibitem[2003]{done}
Done, C., \& Gierli\'nski, M. 2003, \textit{MNRAS}, 342, 1041
\bibitem[2000]{radio}
Fender, R. P., \& Hendry, M. A. 2000, \textit{MNRAS} 317, 1
\bibitem[2002]{gierlinski02}
Gierli\'nski, M., \& Done, C. 2002,  \textit{MNRAS} 331, L47
\bibitem[1989]{hasinger}
Hasinger, C., \& van der Klis, M. 1989,  \textit{A\&A} 225, 79
\bibitem[2001]{lebrun}
Lebrun, F., Leray, J. P., Lavocat, P., et al. 2003, \textit{A\&A} 411, 141
\bibitem[2001]{liu}
Liu, Q. Z., van Paradijs, J., \& van den Heuvel, E. P. J. 2001, \textit{A\&A} 368, 1021
\bibitem[2003]{lund}
Lund, N., Brandt, S., Budtz-Jorgensen, C., et al. 2003,  \textit{A\&A} 411, L231
\bibitem[2003]{miguel}
Mas-Hesse, J. M., Gimenez, A., Culbane, L., et al. 2003,  \textit{A\&A} 411, L261
\bibitem[1984]{mitsuda1}
Mitsuda, K., Inoue, H., Koyama, K., et al. 1984, \textit{PASJ} 36, 741
\bibitem[1989]{mitsuda2}
Mitsuda, K., Inoue, H., Nakamura, N., et al. 1989, \textit{PASJ} 41, 97
\bibitem[2002]{muno}
Muno, M. P., Remillard, R. A., Chakrabarty, D. 2002,  \textit{ApJ} 568, 35
\bibitem[2003]{paizis}
Paizis, A., Beckmann, V., Courvoisier, T.J.-L., et al. 2003, \textit{A\&A} 411, L363
\bibitem[2000]{parmar}
Parmar, A. N., Oosterbroek, T., Del Sordo, S., et al. 2000, \textit{A\&A} 365, 175 
\bibitem[2004]{shaw}
Shaw, S. E., Westmore, M. J., Hill, A. B., et al. 2004, \textit{A\&A} in press, astro-ph/0402587 
\bibitem[2003]{ubertini}
Ubertini, P., Lebrun, F., Di Cocco, G., et al. 2003, \textit{A\&A} 411, L131
\bibitem[1991]{klis2}
van der Klis, M., Kitamoto, S., Tsunemi, H., et al. 1991, \textit{MNRAS}, 248, 751
\bibitem[1995]{klis}
van der Klis, M. 1995 in: \textit{X-Ray Binaries}, Cambridge University Press, 126
\bibitem[2003]{verdr}
Vedrenne, G., Roques, J.-P., Sch\"onfelder, V., et al. 2003,  \textit{A\&A} 411, L63
\bibitem[1986]{white}
White, N. E., Peacock, A., \& Hasinger, G. 1986, \textit{MNRAS} 281, 129
\bibitem[1988]{white2}
White, N. E., Stella, L., \& Parmar, A. N. 1988, \textit{ApJ} 324, 363
\bibitem[2003]{winkler1}
Winkler, C., Courvoisier, T.J.-L., Di Cocco, G., et al. 2003,  \textit{A\&A} 411, L1
\bibitem[2003]{winkler}
Winkler, C., Gehrels, N., Sch\"onfelder, V., et al. 2003b,  \textit{A\&A} 411, L349











\end{thebibliography}
\end{document}